\documentclass[aps,prl,superscriptaddress,twocolumn,twoside,a4paper,floatfix]{revtex4}

\usepackage{hyperref,graphicx,amsmath,latexsym,amssymb,verbatim,revsymb,color,mathpazo}
\newcommand{\ra}{\rangle}
\newcommand{\la}{\langle}
\newcommand{\Tr}{\textrm{Tr}}
\newcommand{\bra}[1]{\langle #1 |}
\newcommand{\ket}[1]{| #1 \rangle}

\begin{document}

\title{How small can thermal machines be? The smallest possible refrigerator}

\author{Noah Linden} \affiliation{Department of Mathematics, University of Bristol$\text{,}$ University Walk, Bristol BS8 1TW, United Kingdom}
\author{Sandu Popescu} \affiliation{H. H. Wills Physics Laboratory, University of Bristol$\text{,}$ Tyndall Avenue, Bristol, BS8 1TL, United Kingdom }
\author{Paul Skrzypczyk} \affiliation{H. H. Wills Physics Laboratory, University of Bristol$\text{,}$ Tyndall Avenue, Bristol, BS8 1TL, United Kingdom }

\date{\today}
\begin{abstract}

We investigate the fundamental dimensional limits to thermodynamic machines. In particular we show that it is
possible to construct self-contained refrigerators (i.e.~not requiring external sources of work) consisting of
only a small number of qubits and/or qutrits. We present three different models, consisting of two qubits, a
qubit and a qutrit with nearest-neighbour interactions, and a single qutrit respectively. We then investigate
fundamental limits to their performance; in particular we show that it is possible to cool towards absolute
zero.
\end{abstract}

\maketitle

When Sadi Carnot \cite{carnot} set out to study the physics of steam engines -- and in the process established
thermodynamics -- the key to progress was to abstract from real machines to idealised, ``model independent''
machines. He found that although the properties of each machine depend on the details of its construction, the
fundamental limit to their efficiency is independent of such details. But can physics be left out completely?
Here we return to physics and ask about other fundamental limits, specifically, is there a fundamental limit to
their size? And, when they are small, are there additional constraints on their performance? Is there a
complementarity between size and performance? For example, can small machines be constructed that cool
arbitrarily close to absolute zero, or does size impose a fundamental limit?

In the present paper we approach these questions in the framework of quantum mechanics which, importantly,
provides a natural and universal notion of ``size'', namely the dimension of the Hilbert space of the system. It
is this measure of size that we will use here to characterise thermal machines. Obviously, spatial extent, mass, thermal capacity, etc. are all possible size measures, each with their own merits. Our measure has an informational flavour,  motivated by the fundamental connection between information and thermodynamics \cite{I1,I2}.

The study of quantum heat engines is, of course, rather well developed \cite{Book1,Book2,B1,B5,B6}. In particular direct quantum analogues of classical Carnot engines have been extensively studied \cite{RR1,RR5}, as well as Otto cycles \cite{G1,G3,G5} However, these papers have in common the fact that they all use an external source of work and/or control -- e.g.~ precise unitary transformations or macroscopic lasers \cite{Bard01}. In this work however, we are interested in fundamental limits on the size of heat engines, hence we must account for \emph{all} degrees of freedom involved; we cannot allow for sources of external work or control. In other words, we want to study \emph{self contained} heat engines, focussing on refrigerators. Clearly no refrigerator can work without a supply a free energy, therefore all we allow ourselves are two heat baths at differing temperatures $T_r < T_h$.

Apart from the implications for fundamental physics, this work is also relevant to other
fields. In biology, for example, if cooling of the active site of a protein can be
achieved then increased catalysis rates may be possible \cite{BriPop08}. It is intriguing to ask
whether simple mechanisms, such as those we describe here, are used by biological systems. A second
field is nanotechnology, where the benefits from cooling at the atomic scale are
clear.

Here we present three models: a refrigerator made of two qubits, one of a qubit and a qutrit with nearest
neighbour interactions and one of a single qutrit -- arguably the smallest possible. We focus on refrigeration of qubits but discuss more general objects also. Finally we prove that there
is no fundamental limit to how close towards absolute zero small refrigerators can cool.

\section{Model I: Two qubits.}

The first model consists of three qubits, two constitute the
refrigerator, and one is the object to be cooled. It is inspired by algorithmic cooling
\cite{schulman-vazirani}, particularly by the few qubit version in \cite{mor}.

 \subsection*{Functioning principle.} Consider first two qubits, for simplicity taken initially to be immersed in the same bath at room temperature $T_r$, but later two different baths. Qubit 1 is the object to be cooled, while qubit 2 will eventually play the role of the ``spiral'' that
takes heat from qubit 1 and dissipates it into the environment.

The free Hamiltonian for the two qubits is $H_0~=~E_1\Pi^{(1)}+E_2\Pi^{(2)}$, where $\Pi^{(i)} =
\ket{1}_i\bra{1}$ is a projector for qubit $i$, $| 0\ra_i$ are the ground states at zero energy, and $|1\ra_i$ are the 
excited states at energies $E_1$ and $E_2$ respectively, where we take $E_2 > E_1$.

At equilibrium each qubit is in a thermal state  $\tau_i$,
\begin{equation}
    \tau_i=r_ie^{-E_i\Pi^{(i)}/kT_r}\label{thermal state}
\end{equation}
where $r_i=\bigl(1+e^{-{E_i/kT_r}}\bigr)^{-1}$ and $k$ is Boltzmann's constant. Since the qubits do not interact
the total thermal state is simply the direct product state $\rho_{12}=\tau_1\otimes\tau_2.$

A convenient way to represent the thermal state (\ref{thermal state}) is in terms of the probabilities, $r_i$
and $1-r_i$, to find the qubit in the eigenstates $|0\ra$ and $|1\ra$ respectively, $\tau_i=r_i|0\ra\la 0|+
(1-r_i)|1\ra\la1|$.

We will denote by $T^{S}_i$ the steady-state temperature of each qubit when refrigeration is occurring. Cooling
of qubit 1 means reaching a temperature $T^{S}_1<T_r$, corresponding to a larger ground state probability $r^{S}_1>r_1$.

The idea of algorithmic cooling is to increase $r_1^S$ by transferring excitations to the second qubit.
Specifically, the probabilities of the eigenstates  $\ket{10}$ and $\ket{01}$ of $\rho_{12}$ are $(1-r_1)r_2$
and $r_1(1-r_2)$. Since we take $E_2 > E_1$, it follows that $(1-r_1)r_2 >(1-r_2)r_1$, i.e. we are more likely to be in the state $\ket{10}$ where qubit 1 is excited than in $\ket{01}$ where it is in the  ground state. Suppose now that we apply a unitary $U$, which swaps these two states,
\begin{equation}
    \ket{10}\longleftrightarrow\ket{01}\label{flip down}
\end{equation}
while leaving the others unchanged.  After the swap we increase the ground state probability of qubit 1, therefore cooling it, and decrease the ground state probability of qubit 2, thus heating it. Since the qubits are in contact with a thermal bath they will eventually return to the environmental temperature $T_r$ if we do nothing else. However, if we keep repeating $U$, qubits 1 and 2 will reach steady-state temperatures $T^{S}_1<T_r$ and $T^{S}_2>T_r$.

The procedure above however requires external work to be performed, since the energy of
$\ket{01}$ is larger than that of $\ket{10}$. This is provided by some external system that induces the
unitary transformation $U$; for example, in an NMR experiment this could be done via a sequence of pulses of an
external magnetic field.

The idea behind our model is to replace the external work by a different source of free energy: Free energy can be provided whenever a system has access to two thermal baths at different
temperatures. In our model this is accomplished by adding a third qubit, in contact with a thermal bath at a hotter temperature $T_h>T_r$. This qubit plays the role of the ``engine''. Thus our fridge
consists in qubits 2 and 3 (the spiral and engine respectively); qubit 1 is the object to be cooled.

To enable transitions between different states without an input of external energy, we take the engine qubit to
have the energy level spacing $E_3$, such that $E_3 = E_2 - E_1$. With this condition we now have two degenerate
energy eigenstates $\ket{010}$ and $\ket{101}$ which can be swapped without requiring work. The
interchange
\begin{equation}
    \ket{101}\longleftrightarrow\ket{010}\label{flip down total}
\end{equation}
accomplishes on qubits 1 and 2 the transformation \eqref{flip down}.

For transitions between $\ket{010}$ and $\ket{101}$ to occur we introduce an interaction Hamiltonian
\begin{equation}
    \label{e:Hint model1} H_{int}=g\big( \ket{010}\bra{101} + \ket{101}\bra{010}\big)
\end{equation}
with $g$ the interaction strength. Note that the interaction is now via a Hamiltonian and not
a unitary transformation. The distinction is important since a unitary necessarily implies external control -- a
device to implement the correct timing, as well as a means of protecting the system from
other external interactions. By using a Hamiltonian no such external control is required.

Furthermore this interaction is taken to be weak compared to the free Hamiltonian, $E_i
\gg g$. In this regime the interaction will not significantly alter the energy eigenvalues or eigenvectors of
the system (which remain governed by $H_0$) and hence we can meaningfully talk about the temperature of the
individual qubits, since each will remain in the standard thermal form \eqref{thermal state}, with $E_1$ and
$E_2$ the same as in the absence of interaction.

To understand the model first note that the interaction Hamiltonian can swap without impediment the states
$\ket{010}$ and $\ket{101}$, since they are degenerate in energy in-so-far as the free Hamiltonian is concerned.
However, if all qubits were kept at the same temperature, the system would be at equilibrium since the the
probability of the flips \eqref{flip down total} are equal. To drive the transitions in one direction and cool
qubit 1, we place the third qubit in a hotter bath; the probability of $\ket{101}$ becomes larger than that of
$\ket{010}$ and so we enhance the probability of the forward flip in \eqref{flip down total} and diminish that of the backward flip. It is this biasing of the interaction which takes heat from qubit 1 into qubit 2,
creating a refrigerator.

\subsection*{Master Equation.}\label{s:M1 eqns} The simplest way to model each qubit being in contact with a thermal bath is to imagine that with probability density $p_i$ per unit time each qubit is thermalised back to its initial thermal state \eqref{thermal state}. (Note that for qubit 3 the bath
temperature is now $T_h$). Mathematically we model this by the non-unitary evolution $\rho \mapsto
p_i\tau_i {\rm Tr}_i \rho + (1-p_i)\rho$. Here $p_i$ quantifies how well insulated each particle is relative to the bath.

All together this leads to the master equation
\begin{equation}
    \label{eqn-of-motion1} \frac{
    \partial \rho}{
    \partial t} = -i[H_0+H_{int}, \rho] + \sum_{i=1}^3 p_i(\tau_i {\rm Tr}_i \rho - \rho),
\end{equation}
(see Appendix), where $H_0=E_1\Pi^{(1)}+E_2\Pi^{(2)}+E_3\Pi^{(3)}$. Note that in general the addition of $H_{int}$ in \eqref{eqn-of-motion1} requires a modification of the dissipative term  if it is to remain consistent \cite{T1}. However, we are interested only in the limit of vanishing $g$ and $p$ such that $g/p =$ constant. In this limit corrections to \eqref{eqn-of-motion1}, of order $pg$, vanish, and the master equation is consistent.

\begin{figure}[t]
    \includegraphics[width=0.9\columnwidth]{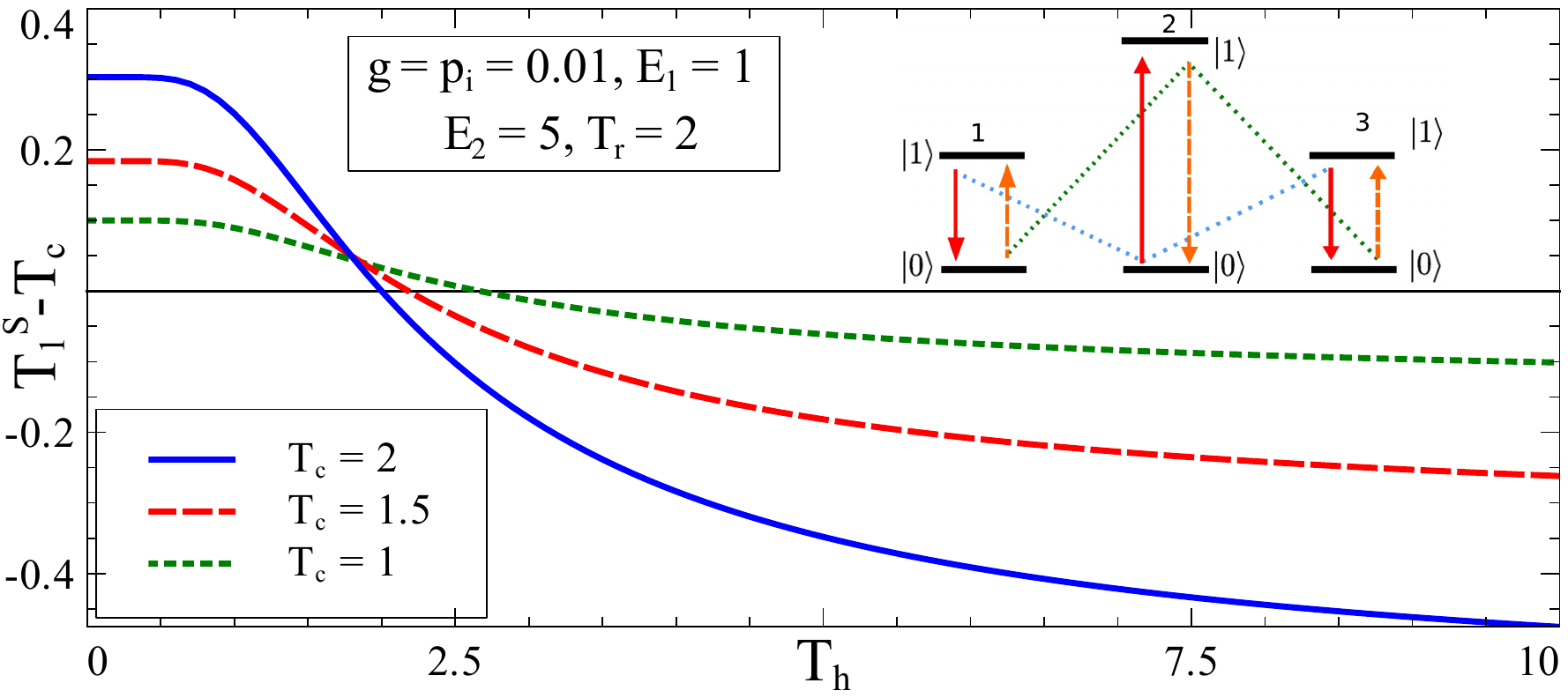} \caption{\label{f:qutrit fridge} Cold qubit steady-state temperature difference $T_1^{S}-T_c$ versus hot bath temperature $T_h$, for various values of $T_c$. Inset: Schematic diagram of energy levels and interaction.}
 	    \includegraphics[width=0.9\columnwidth]{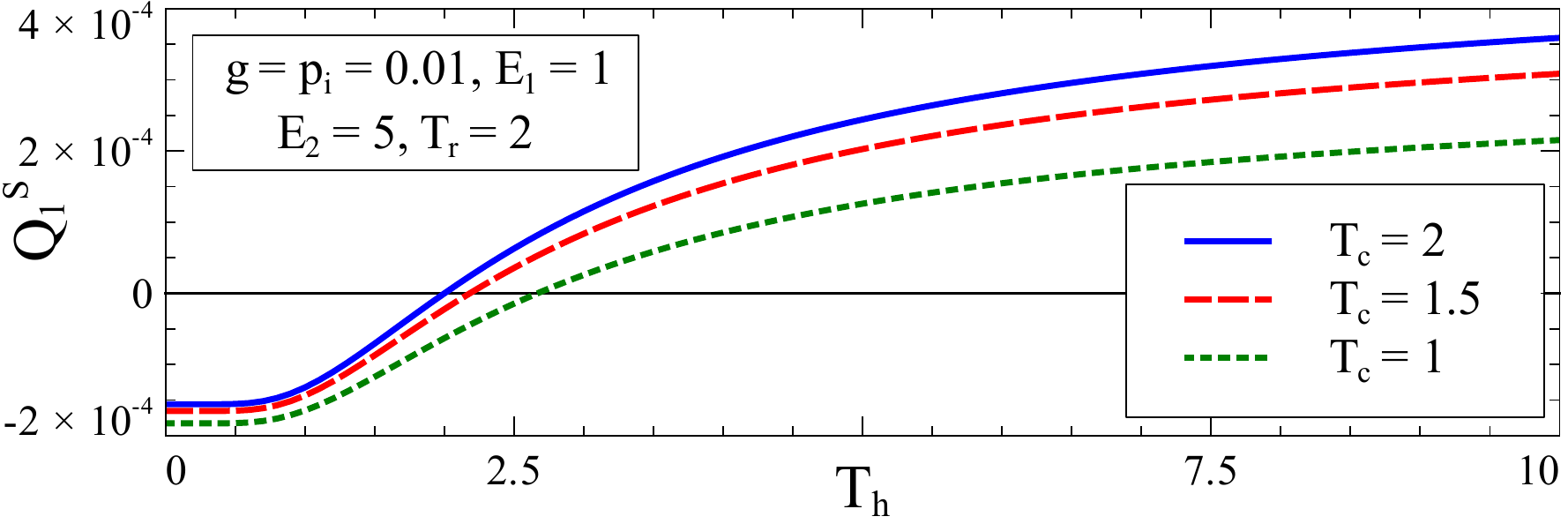} \caption{\label{f:energy} Stationary heat current $Q_1^S$ of qubit 1 versus hot bath temperature $T_h$. $Q_1^S$ becomes positive the moment the cold qubit reaches a temperature colder than its bath.}
\end{figure}

\subsection*{Steady-state Solution.} We will be interested in solving for the stationary state, $\rho_{S}$, which
satisfies $0 = -i[H_{0}+H_{int}, \rho^{S}] + \sum_{i=1}^3 p_i(\tau_i {\rm Tr}_i \rho^{S} - \rho^{S})$. The solution can easily be found analytically, but has a complicated dependence on all the
parameters, therefore it is much more illuminating to present a numerical analysis.

Figure~\ref{f:qutrit fridge} shows the temperature difference between qubit~1 (the system to be cooled) and it's bath, as a function of the temperature of the hot bath, $T_h$. The solid curve shows that when $T_h > T_r$ we are able to achieve cooling, i.e. $T^S_1 < T_r$, and when $T_h = T_r$, that is when we supply no free energy, the temperature of qubit 1 is unchanged. In the dashed and dotted curve we show furthermore that if the cold qubit is in contact with a \emph{colder} bath (at temperature $T_c < T_r$), then we are still able to achieve cooling, i.e. $T_1^S < T_c$, provided $T_h$ is hot enough. Therefore in all instances we demonstrate that our system acts as a refrigerator.

Until now we have focused solely on the task of cooling a qubit. One question is whether our fridge is able to cool more arbitrary objects than just a qubit. To achieve this, consider qubit 1 now as part of the fridge and use \emph{it} to cool other objects. To see that is a valid viewpoint, in Fig.~\ref{f:energy} we display the stationary heat current $Q_1^S$ flowing between the bath and qubit 1. The heat current is defined as $Q_1^S = \Tr(H_1 \mathcal{D}_1(\rho^S))$,\cite{Breuer} where $\mathcal{D}_1(\rho^S) = p_1(\tau_1\otimes\Tr_1\rho^S-\rho^S)$ is the dissipator for qubit 1. We see that a positive current flows whenever the cold qubit is cooled below its bath temperature (c.f.~Fig.~\ref{f:qutrit fridge}). Thus viewing the environment as the arbitrary object, we see that we are able to extract heat from it and therefore cool it. This may also be seen as independent confirmation that our system works as a refrigerator. 

\begin{figure}[b]
	\includegraphics[width=0.9\columnwidth]{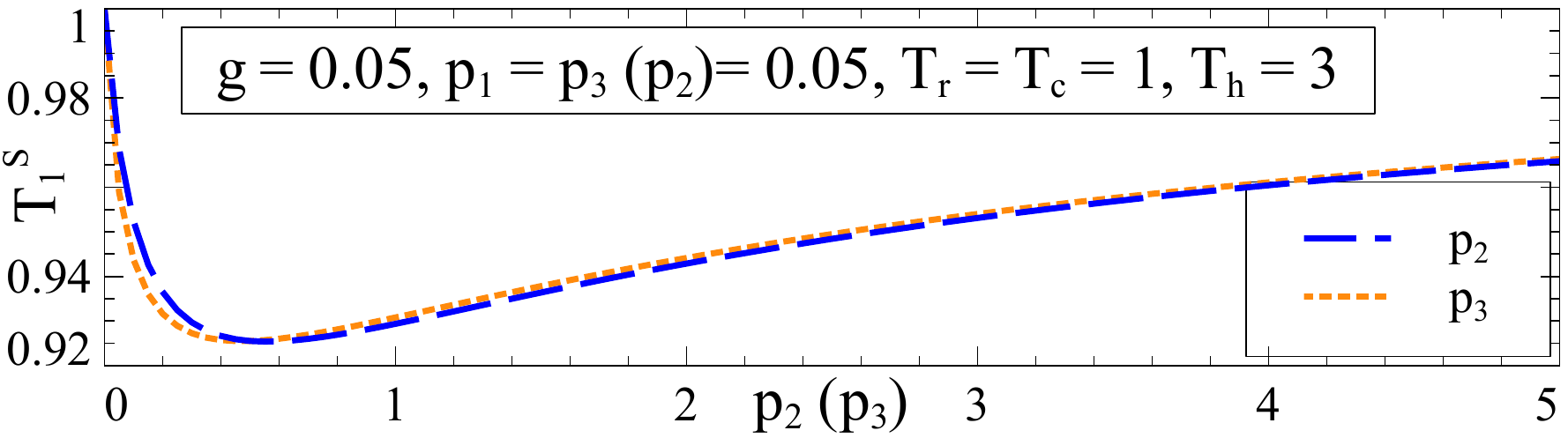} \caption{\label{f:vary p1p2}Qubit 1 steady-state temperature $T_1^{S}$ against insulation parameter $p_2$ ($p_3$) in dashed (dotted) line. When $p_2$ ($p_3$) vanish we are unable to cool. For large $p_2$ ($p_3$) the performance of the fridge degrades due to a Zeno effect.}
		\includegraphics[width=0.9\columnwidth]{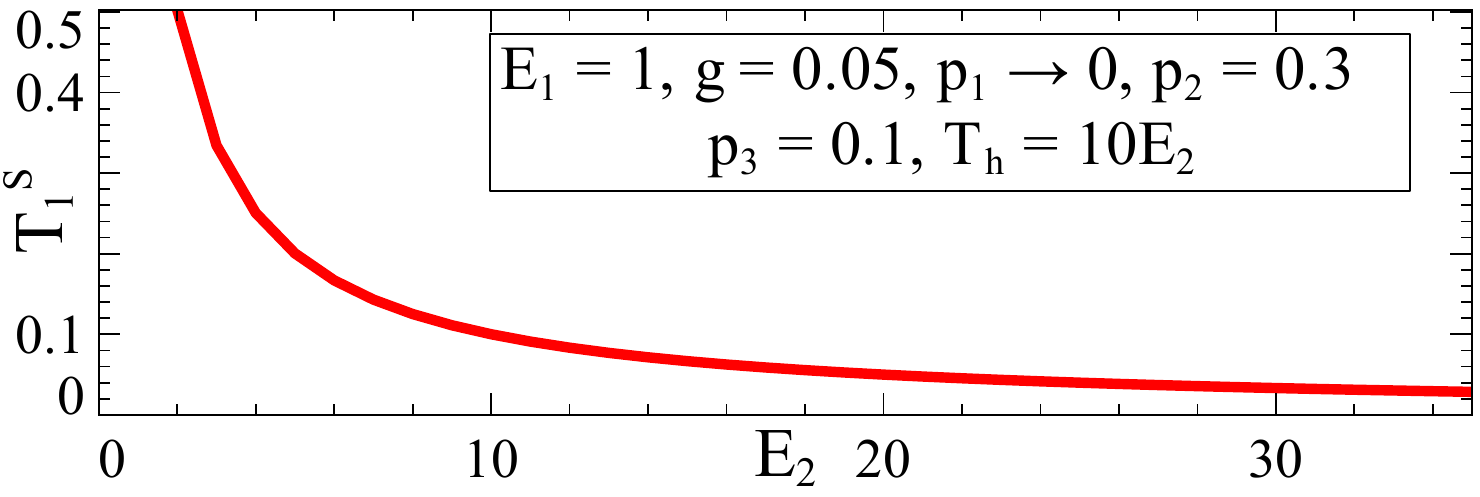} \caption{\label{f:abszero}Qubit 1 steady-state temperature $T_1^{S}$  against energy level spacing $E_2$. As the spacing increases the spin is cooled arbitrarily close to absolute zero.}
\end{figure}

\subsection*{Parameter dependence and Zeno effects.} A natural question to ask is how the behaviour of the fridge
changes as we vary the parameters $p_i$ independently. Indeed, in a `standard' refrigerator we do not want all
parts to interact with the environment equally: the inside of the fridge has to be well insulated to maintain a
low temperature while the spiral at the back of the fridge has to interact strongly with the environment to dissipate heat quickly.


Hence, for qubit 2, (the spiral) we expect that as $p_2$ becomes larger the performance of the
fridge should increase ($T_1^{S}$ should decrease). Furthermore, qubit 3 plays the role of the ``engine'' of
the refrigerator, which it achieves by pumping heat from the hot environment into the system. We thus expect that the
best performance is achieved when it interacts strongly with its environment, as this allows it to
extract heat at the highest rate.

In Fig.~\ref{f:vary p1p2} we plot the dependence of $T_1^{S}$ on $p_2$ and $p_3$. For small values
of $p_2$ and $p_3$ we indeed observe the expected behaviour, however as we increase them further the performance
degrades. The reason is that the quantum Zeno effect \cite{zeno} comes into play in the regime of strong
coupling between the qubits and the environment. Thermalisation is as though the environment measures each qubit. As we increase the rate of thermalisation we enter a regime where the interaction $H_{int}$ will not have time to work between successive thermalisations and hence the refrigerator is no longer able to function.

The dependence of $T_1^S$ on $p_1$ is as expected; coldest temperatures are achieved in the limit of perfect insulation. See Appendix for further details.

\subsection*{Approaching absolute zero.} An important question is whether or not there are \emph{fundamental} limitations on the temperature to which we are able to cool the cold qubit. We show that no such limitations exist.

The minimal achievable temperature is limited by two effects: heat flowing into the fridge due to
imperfect insulation, and the actual cooling ability (i.e. the ability to cool given perfect insulation). It is
the second aspect which we are interested in.


We fix $E_1$, a characteristic of the object to be cooled and not of the refrigerator and also fix
$T_r$, the environmental temperature. We increase $E_3$ (therefore $E_2$ also) and $T_h$ such that the ratio $E_3/T_h$
remains constant and much less than 1. This results in increasing the ground state probability of qubit
2, while maintaining a large excited state probability for qubit 3. Altogether this 
means that the interaction \eqref{e:Hint model1} becomes ever more biased as we increase $E_3$. This leads to
cooling as close as we want towards absolute zero, as seen in Fig.~\ref{f:abszero}.

\section{Model II: One qubit, one qutrit.} One drawback of the previous model is that the interaction Hamiltonian
\eqref{e:Hint model1} is a three body interaction. Here we present a model with only two-body nearest neighbour
interactions.

The model consists of three particles, where one is to be cooled and two construct the fridge. Particles $1$
and $3$ are qubits and particle $2$ is a qutrit. The energy levels of each particle are such that the energy
eigenstates $\ket{020}$ and $\ket{101}$ are degenerate. By introducing an interaction which can take the
population of the latter into the former we can cool down qubit 1.

We do this by introducing two separate interactions between the particles via the Hamiltonians
\begin{eqnarray}
    H_{int}^{(12)} &=& g\big(\ket{02}\bra{11}+\ket{11}\bra{02}\big)\otimes\openone^{(3)}\label{e:Hint trit 1} \\
    H_{int}^{(23)} &=& h\openone^{(1)}\otimes\big(\ket{01}\bra{10}+\ket{10}\bra{01}\big)\label{e:Hint trit 2}
\end{eqnarray}
Neither \eqref{e:Hint trit 1} nor \eqref{e:Hint trit 2} induces transitions between the two desired states.
However, \eqref{e:Hint trit 1} causes transitions between $\ket{020}$ and $\ket{110}$ and \eqref{e:Hint trit 2}
between $\ket{110}$ and $\ket{101}$. Therefore, in \emph{second} order we induce the desired transition. Finally
we bias this interaction, as previously, by taking particles 1 and 2 to be in contact with a bath at temperature
$T_r$ and qubit 3 at $T_h$. This model behaves qualitatively the same as our previous one; details are given in the Supplementary Information.

\section{The smallest possible fridge.} 
\begin{figure}[h] 
 	\includegraphics[width=0.6\columnwidth]{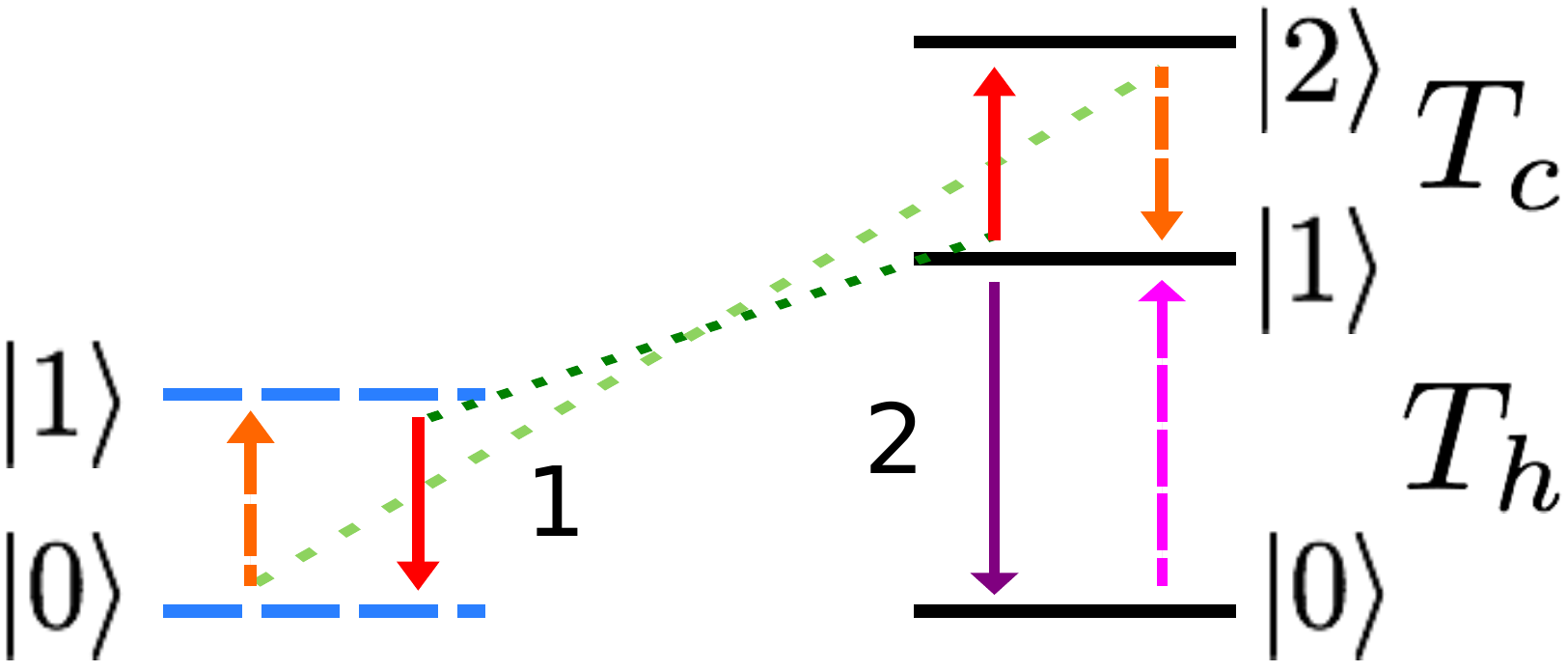} 
	\caption{\label{f:smallest}Schematic diagram of a fridge consisting of a single qutrit (particle 2) with the object to be cooled (particle 1)} 
\end{figure}

In the previous model the qutrit was taken to be in contact with a bath at
temperature $T_r$. However, we could conceive of situations where each of its energy eigenstates has a different
spatial distribution and so can be in contact with environments at differing temperatures. In such situations a
smaller refrigerator can be constructed, by discarding the third qubit -- the fridge now contains only a single
qutrit, see Fig.~\ref{f:smallest}. We believe this is the smallest possible system which may be called a refrigerator.

\section{Conclusions.}~We presented three simple models demonstrating that there is no fundamental
difficulty in constructing small, self contained refrigerators. Moreover showed that it is possible
to cool towards absolute zero.

There are many interesting questions for the future. The first is what can be said about the efficiency of small refrigerators. Is our construction the most efficient or are there other Hamiltonians which are
better for cooling?

Moreover, it is fundamental to ask whether or not there exists a complementarity between small dimension and
efficiency -- can you only be large and efficient or small and inefficient? I.e.
can a small machine reach the efficiency of an ideal Carnot engine? Our particular models do not reach this
efficiency: both the spiral and the engine qubits reach stationary temperatures that differ by a
finite (instead of infinitesimal) amount from the temperatures of their environments, which leads to
irreversible heat exchanges. However, is there a better model? This is not clear, since in a
Carnot cycle the system transitions through infinitely many states, not a finite number as we do here. The question is whether or not this affects the
achievable efficiency of the refrigerator.

Finally it would be interesting to study other thermal machines, for example ones that produce `work'.

\section{Appendix}
\subsection{Deriving the Master equation} To find the master equation consider a small time interval $\delta t$ around a time $t_0$. To first order in $\delta t$, the evolution of the density matrix $\rho$ is given by 
\begin{eqnarray}
	\rho(t_0+\delta t) &=& \big(1-\delta t(p_1+p_2+p_3)\big)\rho(t_0) \nonumber\\
	&+&\delta t\big(p_1\tau_1{\rm Tr}_1\rho(t_0) + p_2\tau_2{\rm Tr}_2\rho(t_0) \\&+& p_3 \tau_3{\rm Tr}_3\rho(t_0)\big) - i \delta t [H_0+H_{int},\rho(t_0)].\nonumber 
\end{eqnarray}
from which it follows that 
\begin{equation}\label{eqn-of-motion1}
	\frac{\partial \rho}{\partial t} = -i[H_0+H_{int}, \rho] + \sum_{i=1}^3 p_i(\tau_i {\rm Tr}_i \rho - \rho), 
\end{equation}
Note that \eqref{eqn-of-motion1} can easily be rewritten explicitly in Lindblad form.

\subsection{Perfect insulation.} In the main text we highlighted specifically the behaviour of the refrigerator as we vary the interaction of qubits 2 and 3 with their respective environments, highlighting Zeno-type effects which arise. Another interesting direction, which we considered when approaching absolute zero, is that of perfect insulation of the cold qubit from its bath. This regime is important as this is where we achieve the optimum performance of the refrigerator, so it is crucial to show that a solution exists in this limit.

\begin{figure}[h] 
	\includegraphics[width=0.9\columnwidth]{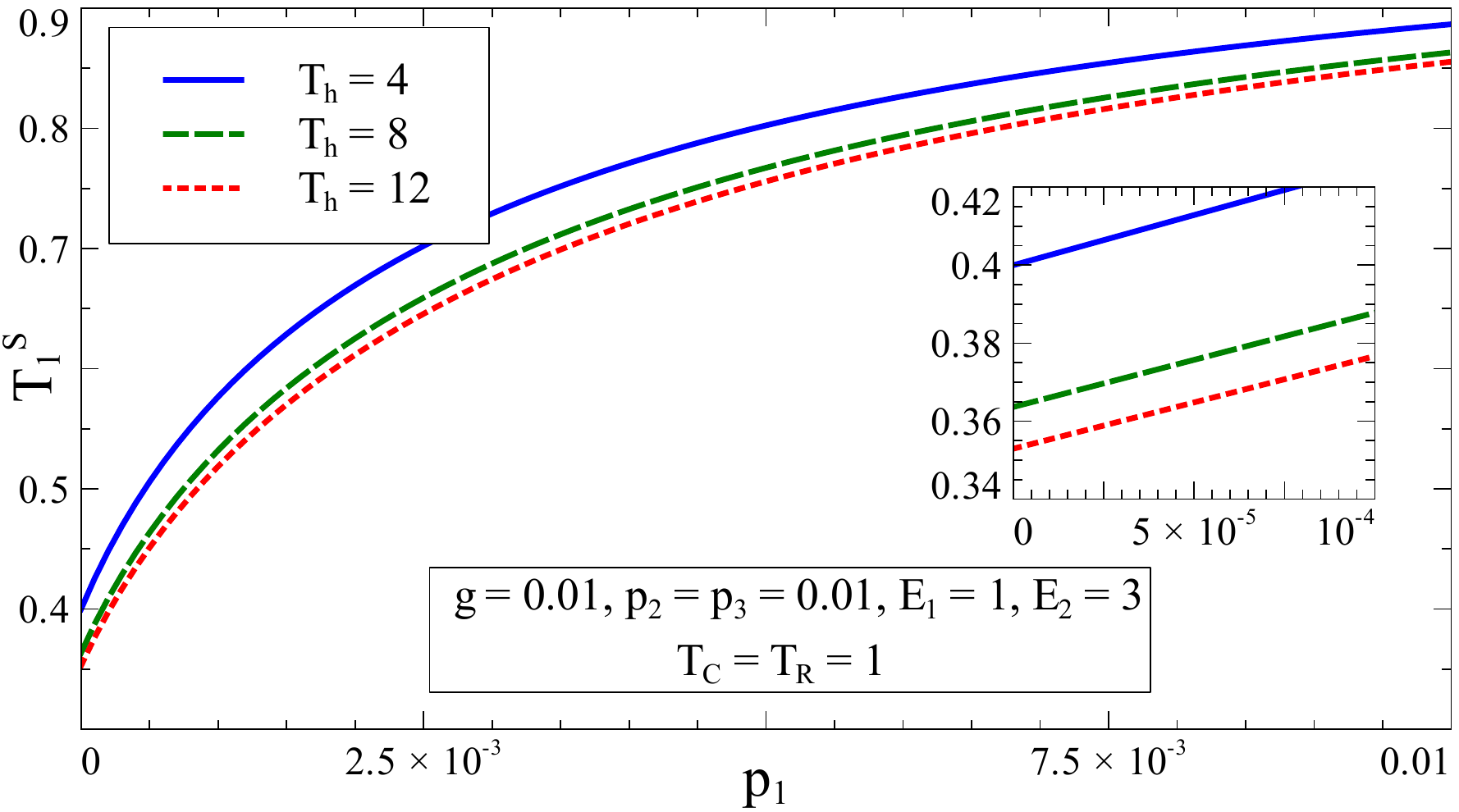} 
	\caption{\label{f:vary p1} Stationary temperature of qubit 1, $T_1^S$ against the insulation parameter $p_1$. For $p_1 \to 0$ we achieve optimal cooling, which we approach continuously from above. (Inset) The limit values, as calculated by Eq. (C1) are 0.400, 0.364, 0.353. } 
\end{figure}

In Fig.~\ref{f:vary p1} we hold all parameters fixed except for $p_1$ and show the dependence of the stationary temperature, $T_1^S$ upon the insulation parameter $p_1$ as we approach perfect insulation, that of $p_1 \to 0$.
	
We observe, for 3 different temperatures of the hot bath, that the curves monotonically increase from the lowest value when $p_1 = 0$. We thus see that as we increasingly insulate the cold qubit it indeed reaches a temperature which remains cold in the limit, as we would expect.

In the paper we have been concerned primarily with studying the behaviour of a fridge numerically, as in general the analytic solution contains a complicated dependence upon the parameters of the model. However, we find that in the limit of perfect insulation that the solution simplifies dramatically and we are able to give a concise expression for the stationary temperature of qubit 1. We find that
\begin{equation}
	T_1^S = \frac{T_C}{1+\tfrac{E_3}{E_1}(1-\tfrac{T_C}{T_h})}
\end{equation}
which is valid as long as $g, p_2, p_3 \neq 0$, and interestingly, in this limit all dependence upon these parameters drops. This expression demonstrates clearly that whenever $T_c < T_h$ that the stationary temperature of qubit 1 is lower than $T_c$, and higher in the opposite case. Furthermore, we see that a lower temperature is achieved by taking the ratio $E_3/E_1$ of energy levels to be large. Using this expression it can easily be checked that for the specific parameters chosen in plotting Fig.~\ref{f:vary p1} that the correct limiting temperature is achieved.

\subsection{Details for two-qubit-one-qutrit refrigerator} The free Hamiltonian of the 3 qubits is taken to be 
\begin{figure}
	\includegraphics[width=0.9\columnwidth]{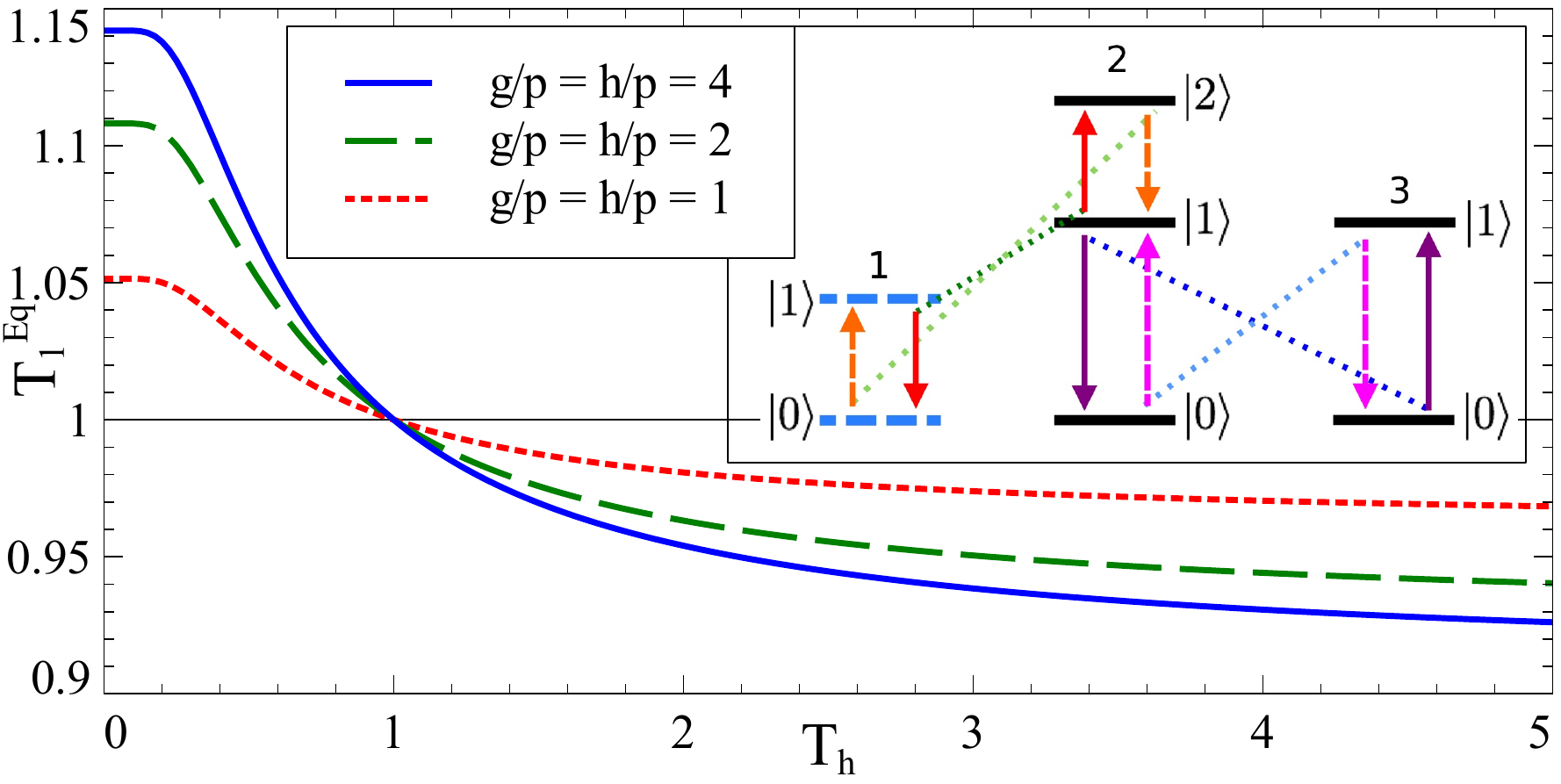} 
	\caption{\label{f:trit temp} Qubit 1 stationary temperature $T_1^{S}$ against qubit 3 bath temperature $T_h$. Here $p_1 = p_2 = p_3 = p$, $T_c = 1$ and $E = 1$. Inset: Schematic diagram showing energy levels and interaction.} 
\end{figure}
\begin{equation}
	H_0 = E_1\Pi_1^{(1)} +E_2\Pi_1^{(2)}+ (E_1+E_2)\Pi_2^{(2)} +E_2\Pi_1^{(3)} 
\end{equation}
where $\Pi_1^{(i)} = \ket{1}_i\bra{1}$ and $\Pi_2^{(i)} = \ket{2}_i\bra{2}$. The interaction Hamiltonian is now 
\begin{eqnarray}\label{e:Hint trit} 
	H_{int} &=& g\big(\ket{02}\bra{11} + \ket{11}\bra{02}\big)\otimes\openone^{(3)}\nonumber \\
	&+& h\openone^{(1)}\otimes\big(\ket{01}\bra{10} + \ket{10}\bra{01}\big) 
\end{eqnarray}
where $g$ and $h$ are the coupling constants for each interaction. We again assume that the interaction strength between the qubits is small in comparison to the energy level spacing, $E_i$, that is $g,h \ll E_i$. Each particle again interacts with a thermal bath. For qubits $1$ and $3$, the thermal states are given, as in the main text, Equation (2). For particle $2$ the thermal state is now given by 
\begin{equation}
	\tau_2 = N'_2\exp(-(E_2\Pi_1^{(2)}+(E_1+E_2)\Pi_2^{(2)})/kT_c) 
\end{equation}
where $N'_2 = (1+e^{-E_1/kT_c}+e^{-(E_1+E_2)/kT_c})^{-1}$.

The master equation governing the evolution is 
\begin{equation}
	\frac{\partial \rho}{\partial t} = -i[H_0+H_{int}, \rho] + \sum_{i=1}^3 p_i(\tau_i {\rm Tr}_i \rho - \rho) 
\end{equation}
for which we will be interested in solving for the stationary solution $\rho_{S}$. As before, we will not present an analytic form for $\rho_{S}$ but only study properties of it numerically. Figure \ref{f:trit temp} displays the dependence of the steady-state temperature of qubit 1, $T_1^{S}$ on $T_h$, the temperature of the hot bath, for various values of the parameters $g/p$ and $h/p$. We observe that this refrigerator behaves qualitatively the same as in our previous model.

\end{document}